\documentclass[prd,preprintnumbers,superscriptaddress,nofootinbib,showpacs]{revtex4-1}
\usepackage{hyperref}
\usepackage{pstricks}
\usepackage{color}
\usepackage{amssymb,amsmath,bm}
\usepackage{epsfig}
\usepackage{afterpage}
\usepackage{longtable}
\usepackage{latexsym, mathrsfs}
\usepackage{graphics}
\usepackage{url}
\usepackage{epstopdf}
\usepackage{subfigure}
\usepackage{tikz,pgfplots}
\usetikzlibrary{shapes,arrows}
\usepackage{tabls}

\begin{document}
\preprint{BARI-TH/719-19}

\title{Heavy pentaquark spectroscopy in the diquark model}
\author{Floriana~Giannuzzi}
\affiliation{Universit\`a degli Studi di Bari,via Orabona 4, I-70126 Bari, Italy}

\begin{abstract}
$QQ^\prime qq\bar q$ pentaquarks are studied in a potential model, under the hypothesis that they are composite objects of two diquarks and one antiquark. The interaction between two colored objects includes two contributions, one based on the $q\bar q$ potential in QCD, computed in the gauge/string duality
approach, and another describing the spin-spin interaction. The model has been extended to investigate pentaquarks with different quark content, as  $Qqqq\bar q$ and $Qqqq\bar Q$, the latter including the states observed by LHCb, $P_c(4380)^+$ and $P_c(4450)^+$, later updated,  with a new data sample, to $P_c(4312)^+$, $P_c(4440)^+$, and $P_c(4457)^+$. 
\end{abstract}

\vspace*{2cm}
%\pacs{12.39.Ki, 12.39.Pn, 14.20.Lq, 14.20.Mr} 

\maketitle

\section{Introduction}
Quantum chromodynamics (QCD) foresees the existence of quark-antiquark states (mesons) and three-quark states (baryons), as well as multiquark states \cite{GellMann:1964nj}-\cite{Jaffe:2004ph} such as tetraquarks, comprising two quarks and two antiquarks, and pentaquarks, comprising four quarks and one antiquark. The only requirement for these states is to be color singlets. Although most of the ground state mesons and baryons are experimentally well known, many recently observed states are under discussion since their quark content and/or spin/parity are uncertain \cite{Tanabashi:2018oca}-\cite{Colangelo:2012xi}; for a review on possible exotic states see \cite{Ali:2017jda}-\cite{Karliner:2017qhf}-\cite{Chen:2016qju}.
One of the most intriguing cases is the $X(3872)$, first observed by the Belle Collaboration \cite{Choi:2003ue}. The spin parity assignment $1^{++}$ is compatible with a meson state in the quark model  \cite{Colangelo:2007ph}, however its decay channels suggested a possible interpretation as a four-quark state \cite{Maiani:2004vq}-\cite{Voloshin:2003nt}.    
In 2015, LHCb observed two resonances in the $J/\psi \, p$ channel in $\Lambda_b^0$ decay, labeled $P_c^+$, with mass $4380\pm 8\pm 29$ MeV and $4449.8\pm 1.7\pm 2.5$ MeV, opposite parity and spin $3/2$ and $5/2$, compatible with heavy pentaquark $c\bar c u u d$ states  \cite{Aaij:2015tga}. Later on, in 2019, during the \emph{Rencontres de Moriond} conference, the LHCb Collaboration announced  \cite{newlhcb} the observation,  in the same energy region, of  the resonances $P_c(4312)^+$, $P_c(4440)^+$, and $P_c(4457)^+$ \cite{Aaij:2019vzc}. According to this new analysis, the previously reported state $P_c(4450)^+$ contains two narrow peaks, corresponding to $P_c(4440)^+$ and $P_c(4457)^+$.
Previously, in 2003 researchers from the SPring-8 laboratory in Japan \cite{Nakano:2003qx}, ITEP in Russia \cite{Barmin:2003vv}, Jefferson Lab in Virginia \cite{Stepanyan:2003qr}, and from the ELSA accelerator in Germany \cite{Barth:2003es} announced the observation of the $\Theta^+$ pentaquark, consisting of four light quarks and a strange antiquark, but such evidence has not been confirmed by later experiments \cite{Hicks:2012zz}.

Considering that  many experimental results will be allowed in the next few years by the increasing luminosity at experiments like LHCb at CERN and Belle-II at SuperKEKB, in this paper we compute the masses of heavy pentaquarks using a potential model.
Our results can be compared with outcomes of different studies that appeared in the past few years in this sector. For example, the masses of $Q\bar Q qqq$, $Q$ being a heavy quark and $q$ a light quark, have been computed in  \cite{Deng:2016rus}-\cite{Azizi:2017bgs}, while \cite{Hiyama:2018ukv}-\cite{Richard:2017una}-\cite{Ali:2016dkf}-\cite{Azizi:2016dhy} focus on the hidden-charm $c\bar c qqq$ pentaquarks, having the same quark content as the states observed by LHCb. In \cite{Santopinto:2016pkp} a classification of all possible $Q\bar Q qqq$ states and quantum numbers has been presented. $Q Q qq\bar q$ states have been considered in, e.g., \cite{Wang:2018lhz}-\cite{Zhou:2018pcv}.

In the investigation of multiquark states, one of the most discussed issues is the existence of possible internal structures. In this respect, the main hypotheses for pentaquarks are that they could be compact states, relying on the interaction among two diquarks and an antiquark \cite{Maiani:2015vwa}, or molecular states, relying on the interaction between a baryon and a meson \cite{Karliner:2015ina}-\cite{Chen:2015loa}. 
Following the former approach, in Section \ref{sec:QQ} we  compute the masses of $QQ^\prime qq\bar q$ pentaquarks using the model introduced in Section \ref{sec:model}.
An attempt to study pentaquarks with a different quark content is put forward in Sections \ref{sec:Qq}-\ref{sec:QQbar}. Section \ref{sec:conclusions} contains discussions and conclusions.

\section{Model}\label{sec:model}
We study pentaquarks in the potential model introduced in \cite{Carlucci:2007um}, in which meson masses are computed by solving the wave equation:
\begin{equation}\label{salpetereq}
\left(\sqrt{m_1^2-\nabla^2}+\sqrt{m_{2}^2-\nabla^2}+{V}(r)\right)
\psi({\bf r})\,=\,M\, \psi({\bf r})\, ,
\end{equation} 
where $m_1$ and $m_2$ are the masses of the constituent quark and antiquark, $V(r)$ is the quark-antiquark potential, $M$ and $\psi$ are the mass and wave function of the meson. Eq.~\eqref{salpetereq} arises from the Bethe-Salpeter equation in QCD by considering an instantaneous local potential of interaction. Differently from the Schr\"odinger equation, it has relativistic kinematics.

\begin{figure}[h!]
\begin{center}
    \includegraphics[width=11.5cm]{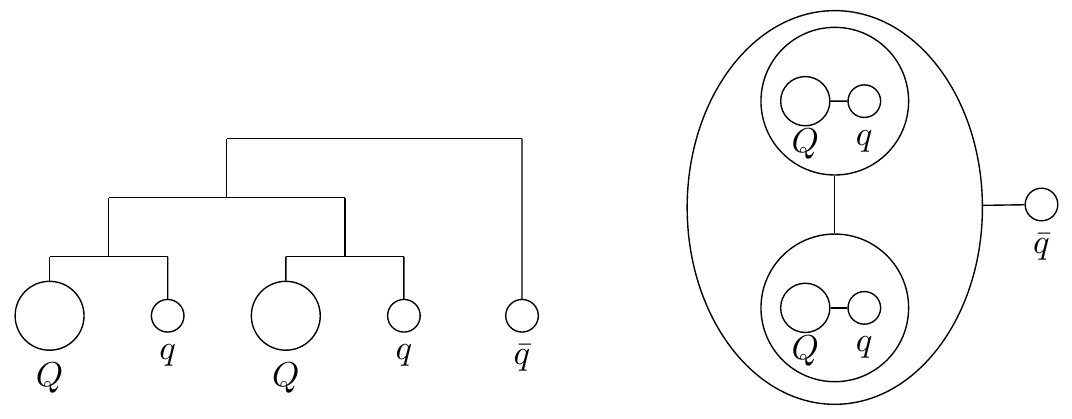}
\caption{Dendrogram and picture of the quark content of the pentaquark in the diquark-diquark-antiquark model (model $\mathcal{A}$).}\label{figuraddq}
\end{center}
\end{figure}

Eq.~\eqref{salpetereq} can be also used  to study pentaquarks if a pentaquark is considered as the bound state of two diquarks and an antiquark. The strategy consists in computing at first diquark masses and wave functions from interactions between single quarks, then the mass and wave function of the four-quark state formed by two diquarks, and finally the mass and wave function of the pentaquark resulting from the interaction between the four-quark state and one antiquark, as shown in Fig.~\ref{figuraddq}. We call this model $\mathcal{A}$. In this picture, each interaction is between two objects, as in Eq.~\eqref{salpetereq}. 
Model $\mathcal{A}$ is based on the diquark-diquark-antiquark description of pentaquarks   \cite{Maiani:2015vwa}, and on SU(3) color group arguments, according to which two quarks (in the $\mathbf{3}$ representation of the group) can attract each other forming a diquark (in the $\mathbf{\bar3}$ representation), and similarly two diquarks can attract each other forming a four-quark state (in the $\mathbf{3}$ representation of the group), and finally a four-quark state plus an antiquark form a color singlet (the pentaquark). We adopt the one-gluon-exchange approximation, in which the potential of  interaction between two quarks ($\mathbf{3}\otimes\mathbf{3}$) is equal to half the $q \bar q$ potential ($\mathbf{3}\otimes\mathbf{\bar3}$) \cite{Richard:1992uk}.

The $q\bar q$ potential used in \eqref{salpetereq} for each two-body interaction has three terms:
\begin{equation}\label{potenziale}
V(r) = V_{QCD}(r) + V_{spin}(r) + V_0\,,
\end{equation}
where $V_0$ is a constant term, $V_{QCD}(r)$ represents the color interaction and $V_{spin}(r)$ the spin-spin interaction.

\begin{figure}
  \begin{center}
    \includegraphics[width=7cm]{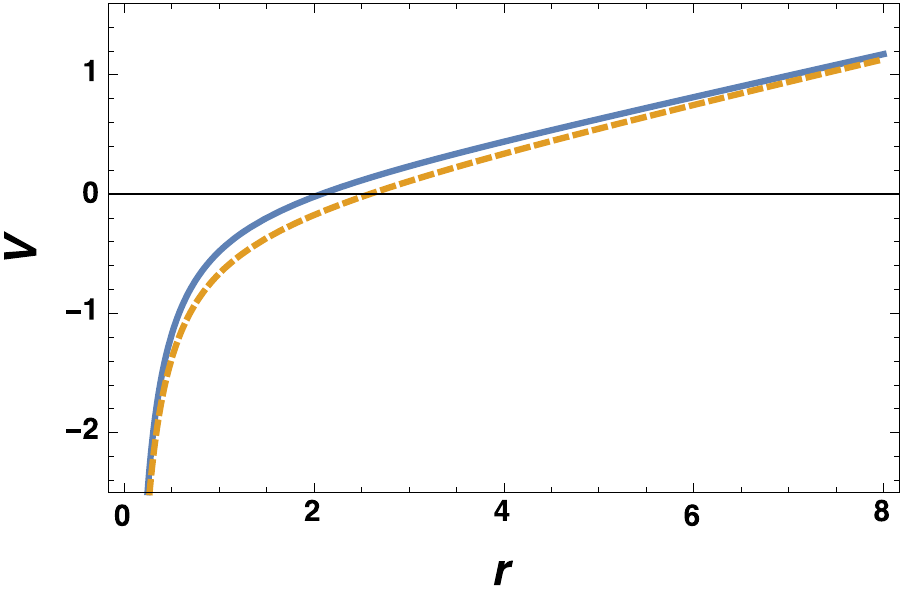}
    \caption{Blue plain line: $V_{QCD}(r)$ from Eq.~\eqref{adsqcdpot} for $c=0.30$ GeV$^2$ and $g=2.75$. Orange dashed line: Cornell potential $V(r)=-\frac{a}{r}+b\, r+C$ for $a=0.63$, $b=0.18$ GeV$^2$ and $C=-0.22$ GeV, fixed from the requirement that the Cornell potential has the same asymptotic behavior as $V_{QCD}(r)$.}
    \label{fig:potential}
  \end{center}
\end{figure}

For $V_{QCD}(r)$ we use the potential found in \cite{Andreev:2006ct} in a phenomenological model inspired by the AdS/QCD correspondence by computing the expectation value of a rectangular Wilson loop. The potential is given in parametric form:
\begin{equation}\label{adsqcdpot}
\left\{
\begin{array}{cc}
\displaystyle
V_{QCD}(\lambda)\,=\,\frac{g}{\pi}
\sqrt{\frac{c}{\lambda}} \left( -1+\int_0^1 dv \, v^{-2} \left[
\mbox{e}^{\lambda v^2/2} \left(1-v^4
\mbox{e}^{\lambda(1-v^2)}\right)^{-1/2}-1\right]\right) & \\
\displaystyle r(\lambda)\,=\,2\, \sqrt{\frac{\lambda}{c}} \int_0^1 dv\, v^{2}
\mbox{e}^{\lambda (1-v^2)/2} \left(1-v^4
\mbox{e}^{\lambda(1-v^2)}\right)^{-1/2}  \hspace{3.3cm}&  \,,
\end{array}
\right.\end{equation}
where $r$ is the  distance between the quark and antiquark, $c$ and $g$ are  parameters. A comparison between $V_{QCD}(r)$ and the Cornell potential $V(r)=-\frac{a}{r}+b\, r+C$ \cite{Eichten:1978tg} is shown in Fig.~\ref{fig:potential},  for values of parameters $c=0.30$ GeV$^2$ and $g=2.75$ in Eq.~\eqref{adsqcdpot}, and $a=0.63$, $b=0.18$ GeV$^2$, $C=-0.22$ GeV for the Cornell potential, these latter ensuring the two potentials have the same asymptotic behavior at large and small distances. One can notice that the Cornell potential is slightly lower than $V_{QCD}(r)$ in the middle-distance region.

The term $V_{spin}(r)$ is given by \cite{Barnes:2005pb}: 
\begin{equation}\label{spinpot}
V_{spin}(r)\,=\,A \frac{\tilde\delta(r)}{m_1 m_2}{\bf S_1}\cdot{\bf
S_2} \qquad\quad\mbox{with }\qquad \tilde\delta(r)=\left(\frac{\sigma}{\sqrt{\pi}}\right)^3
e^{-\sigma^2 r^2}\,, 
\end{equation}
where $\sigma$ is a parameter defining the smeared delta function and  ${\bf S}$ is the spin of the interacting particle.
As usual, we use the trick
\begin{equation}\label{spincombination}
{\bf S_1}\cdot{\bf
S_2} = \frac{1}{2}\left( S(S+1)-S_1(S_1+1)-S_2(S_2+1)\right) \,,
\end{equation}
$S$ being the total spin.
The parameter $A$ is proportional to the strong coupling constant $\alpha_s$ in the one-gluon-exchange approximation.

A cutoff at small distance is introduced to cure the singularity of the wave function, fixing the potential \eqref{potenziale} at the value ${V}(r_M)$ for $r\leqslant r_M$, with $r_M=\frac{k}{M}$ in case $m_1=m_2$, and $r_M=\frac{k^\prime}{M}$ in case $m_1\neq m_2$  \cite{Cea:1986bj}-\cite{Colangelo:1990rv}. $k$ and $k^\prime$ are two parameters and $M$ is the mass of the final state.

Notice that both the one-gluon-exchange approximation and the use of an instantanous potential can be properly applied only to heavy states, in which at least one of the two interacting particles is heavy, i.e. contains a charm or bottom quark. Therefore, we  compute masses of pentaquarks containing at least one heavy quark. Moreover, at each step we only consider states with orbital angular momentum $\ell$=0.

We solve the Salpeter equation \eqref{salpetereq} through the Multhopp method \cite{Colangelo:1990rv}, which allows one to transform an integral equation into a set of linear equations containing  variables called Multhopp's angles.
We fix the parameters of the model as in \cite{Carlucci:2007um}, where the masses of heavy mesons have been fitted to their experimental values:
%
%\begin{table}[h!]
\begin{center}
\begin{tabular}{ccc}
$c=0.300$ GeV$^2$ & $g=2.750$ & $V_0=-0.488$ GeV\\
$A_c=7.920$  & $A_b=3.087$  & $\sigma=1.209$  GeV\\
$m_q=0.302$ GeV & $m_s=0.454$ GeV & $m_c=1.733$ GeV \\
& $m_b=5.139$ GeV . &
\end{tabular}
\end{center}
%\end{table}%
% 
Two values for the parameter $A$ in \eqref{spinpot} have been introduced, in order to take into account the two scales, $\mathcal{O}(m_c)$ and $\mathcal{O}(m_b)$, at which $\alpha_s$ must be computed: $A_b$ is used for states comprising at least a  beauty quark and $A_c$ otherwise.
The model, with this choice of parameters, has been able to predict with very good accuracy the mass of $\eta_b$ \cite{Giannuzzi:2008pv}, observed soon after by the BABAR Collaboration \cite{Aubert:2008vj}.

As a first step, diquark masses are obtained by solving Eq.~\eqref{salpetereq} with potential \eqref{potenziale} divided by a factor 2 and a cutoff at $r=r_M$, as done in \cite{Carlucci:2007um}. 
In the second step, we use the Salpeter equation to study the interaction between two diquarks. The diquark-diquark potential is assumed to be the same as between two quarks in a diquark: this suggests to adopt again the potential \eqref{potenziale} divided by a factor 2. However,  diquarks are extended objects, so we take into account the structure of the diquarks by
defining a smeared potential \cite{Carlucci:2007um}: 
\begin{equation} \label{convpotential}
\tilde V(R)=\frac{1}{N}\int
d{\bf r_1}\int d{\bf r_2}|\psi_d({\bf r_1})|^2|\psi_d({\bf r_2})|^2
V\Big(\Big|{\bf R}+{\bf r_1}-{\bf r_2}\Big|\Big)\,.\displaystyle
\end{equation} 
In this equation $\psi_{d}$ is the diquark wave function, $N$ is a normalization factor. 
Since $|\psi_{d}({\bf r})|^2$ is strongly peaked at
$r\sim 0$, we cut the integral  at the peak value of the function $u_{d}(r)=r \psi_d(r)$  \cite{Carlucci:2007um}.
In the last step, the potential producing a singlet state is obtained from a convolution of \eqref{potenziale} with the diquark-diquark $\psi_{dd}$ wave function  \cite{Giannuzzi:2009gh}:  
\begin{equation}\label{convpotentialddq}
\hat V(R)=\frac1{N^\prime}\int d{\bf r} \; |\psi_{dd}({\bf r})|^2
V( |{\bf R}+{\bf r}|)
\end{equation}
with $N^\prime$ a normalization factor.

\section{Results}

\subsection{$QQ^\prime qq\bar{q}$}\label{sec:QQ}
Using the model $\mathcal{A}$ introduced in the previous Section, we compute the masses of pentaquarks comprising two heavy quarks. Each heavy quark forms a diquark with one light quark, therefore these states can be well described in this framework.
In Table~\ref{tabdiquark} the masses and spin couplings $\kappa$ of heavy diquarks are shown, in which the spin coupling $\kappa$ is defined as the coefficient  multiplying $e^{-\sigma^2 r^2}$ in the spin-spin interaction potential \eqref{spinpot}, i.e. $\kappa=\frac{A}{2} \frac{1}{m_1 m_2} \left(\frac{\sigma}{\sqrt{\pi}}\right)^3 {\bf S_1}\cdot{\bf S_2}$, where the factor $1/2$ is due to the one-gluon exchange approximation for the quark-quark interaction; Table~\ref{tabdiquark} also contains the values of $\bar{\kappa}$ defined as
\begin{equation}\label{kappamean}
\bar{\kappa}=\int d{\bf r} \; \psi_{d}(r)^2 \frac{1}{2} V_{spin}(r) \,,
\end{equation}
with $V_{spin}(r)$ from Eq.~\eqref{spinpot} and $\psi_{d}$ the diquark wave function.
We adopt the following notation: $[Qq]$ diquark has spin 0, and $\{Qq\}$ has spin 1.
\begin{table}[h]
\caption{Masses (GeV) and spin couplings (GeV) of diquarks. $[Qq]$ diquark has spin 0, and $\{Qq\}$ has spin 1. $\kappa=\frac{A}{2} \frac{1}{m_1 m_2} \left(\frac{\sigma}{\sqrt{\pi}}\right)^3 {\bf S_1}\cdot{\bf S_2}$, while $\bar{\kappa}$ is defined in Eq.~\eqref{kappamean}.}
\begin{center}
\begin{tabular}{|c|c|c|c|c|c|c|c|c|}
\hline\hline
 & $[cq]$ & $\{cq\}$  & $[bq]$ & $\{bq\}$  & $[cs]$ & $\{cs\}$  & $[bs]$ & $\{bs\}$ \\ 
 \hline\hline
 Mass & 2.118 & 2.168 & 5.513 & 5.526 & 2.237 & 2.276 & 5.619 & 5.630 \\
 \hline
 $\kappa$ & -1.799 & 0.600 & -0.236 & 0.079 & -1.197 & 0.399 & -0.157 & 0.052 \\
 \hline
  $\bar{\kappa}$ & -0.053 & 0.009 & -0.012 & 0.003 & -0.055 & 0.008 & -0.009 & 0.002 \\
\hline\hline
\end{tabular}
\end{center}
\label{tabdiquark}
\end{table}%

Pentaquark masses are shown in Table \ref{pentaQQ}, where $q=\{u,d\}$, $q^\prime=\{u,d,s\}$ and $Q,Q^\prime=\{c,b\}$. Since we  set $\ell=0$ in all the cases,  the states have negative parity.  
As for tetraquarks \cite{Carlucci:2007um}, a large number of states with different spin is found when combining two diquarks and an antiquark. 
There are five spin-1/2, four spin-3/2 and one spin-5/2 states, as expected when combining five spin-1/2 particles. For the sake of completeness we should mention that not all the states can be considered in the spectrum since one must take spin, flavor and color representations such that the total wavefunction of identical fermions (bosons) is antisymmetric (symmetric). More details about this topic can be found in \cite{Santopinto:2016pkp}-\cite{Zhou:2018pcv}. 

\begin{table}[h]
\caption{Masses (GeV) of $QQ^\prime qq\bar q$ pentaquarks, where $q=u,d$ and $Q,Q^\prime=\{c,b\}$. }
\begin{center}
\begin{tabular}{|c|c|c|c|c|c|c|c|}
\hline\hline
Content & $J^P$& \multicolumn{2}{|c|}{Mass ($Q,Q^\prime=c$)} & \multicolumn{2}{|c|}{Mass ($Q,Q^\prime=b$)}  & \multicolumn{2}{|c|}{Mass ($Q=b,Q^\prime=c$)}\\
\cline{3-8}
& & $q^\prime=u,d$ & $q^\prime=s$& $q^\prime=u,d$ & $q^\prime=s$& $q^\prime=u,d$ & $q^\prime=s$\\
\hline\hline
$\bar q[Qq][Q^\prime q^\prime]$ &  $\frac{1}{2}^-$ & 4.54 & 4.66 & 11.15 & 11.25 & 7.85 & 7.96\\
\hline
$\bar q\{Qq\}[Q^\prime q^\prime]$ &   $\frac{1}{2}^-$ & 4.57 & 4.68  &  11.16 & 11.26 & 7.86 & 7.97\\
\hline
$\bar q[Qq]\{Q^\prime q^\prime\}$ &   $\frac{1}{2}^-$ & 4.57  & 4.66 & 11.16 & 11.25 & 7.92 & 8.01\\
\hline
$\bar q(\{Qq\}\{Q^\prime q^\prime\})_{s=1}$ &   $\frac{1}{2}^-$ & 4.64 & 4.73 & 11.19 & 11.28 & 7.94 & 8.04\\
\hline
$\bar q(\{Qq\}\{Q^\prime q^\prime\})_{s=0}$ &   $\frac{1}{2}^-$ & 4.69 & 4.78 & 11.20 & 11.29 & 7.96 & 8.05 \\
\hline
$\bar q(\{Qq\}\{Q^\prime q^\prime\})_{s=2}$ &   $\frac{3}{2}^-$ &  4.62 & 4.72 & 11.18 & 11.27 & 7.94 & 8.03 \\
\hline
$\bar q\{Qq\}[Q^\prime q^\prime]$ &   $\frac{3}{2}^-$ & 4.65 & 4.77 &  11.18 & 11.28 & 7.89 & 8.00 \\
\hline
$\bar q[Qq]\{Q^\prime q^\prime\}$ &   $\frac{3}{2}^-$ & 4.65 & 4.75 &  11.18 & 11.28 & 7.95 & 8.04 \\
\hline
$\bar q(\{Qq\}\{Q^\prime q^\prime\})_{s=1}$ &   $\frac{3}{2}^-$ &  4.72 & 4.82 & 11.21 & 11.30 & 7.97 & 8.06\\
\hline
$\bar q\{Qq\}\{Q^\prime q^\prime\}$ &   $\frac{5}{2}^-$ & 4.75 & 4.85 &  11.22 & 11.31 & 7.98 & 8.07 \\
\hline\hline
\end{tabular}
\end{center}
\label{pentaQQ}
\end{table}%

\subsection{$Qqqq\bar{q}$}\label{sec:Qq}
\begin{figure}[h!]
\begin{center}
    \includegraphics[width=11.5cm]{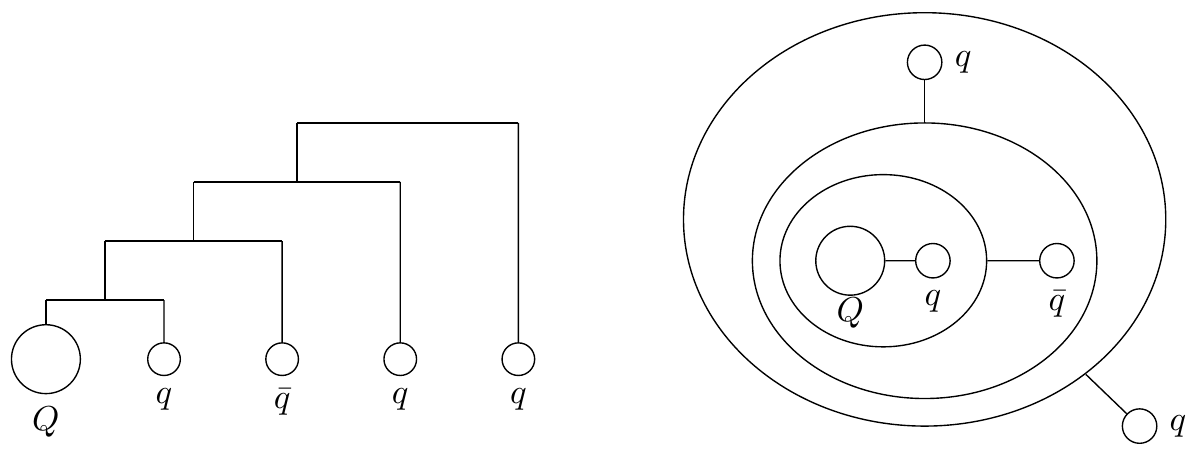}
\caption{Dendrogram and sketch of the quark content of the pentaquark in the  model $\mathcal{B}$ described in Section \ref{sec:Qq}.}\label{figuredqqq}
\end{center}
\end{figure}

Let us consider pentaquarks with one heavy quark and four light quarks. Model $\mathcal{A}$ introduced in Section \ref{sec:model} cannot be used here, since only one heavy diquark can be constructed, while the other diquark would contain two light quarks. Nevertheless, we try to determine the masses of these states by considering the singlet state resulting from subsequent interaction of the heavy quark  with a light quark,  as a sequence of the two-body interactions sketched in Fig.~\ref{figuredqqq}. This model, labeled $\mathcal{B}$, is described hereinafter. Although  such a configuration of quarks inside the pentaquark is not usually expected, it is interesting to investigate this possibility, which can also be applied to {$Qqqq\bar{Q}$} states (see Section \ref{sec:QQbar}), the candidates for the peaks observed by LHCb in the hidden charm sector.

The strategy of the computation is described in the following scheme, in which each step consists in solving Eq.~\eqref{salpetereq} for the indicated particles:
\begin{enumerate}\label{scheme}
\item $Q + q \to Qq$
\item $Qq + \bar q \to Qq\bar q$
\item $Qq\bar q + q \to Qq\bar qq$
\item $Qq\bar qq +  q \to Qqqq\bar q$ .
\end{enumerate}
The first three interactions are between states in the same representation of the color group ($3$ or $\bar 3$), while only the last one produces a color singlet.
We treat the last interaction in the same way as the one between a quark and an antiquark, i.e. by solving Eq.~\eqref{salpetereq} with a $q\bar q$ potential, $m_1$ being the mass of the quark and $m_2$ the mass of the four-quark state. Eq.~\eqref{salpetereq} has been used to study the first three interactions as well, with a potential equal to half the $q \bar q$ potential \cite{Richard:1992uk}. The correction \eqref{convpotentialddq} to the potential has been considered.

The masses of  pentaquarks  with one charm or one beauty are shown in Table \ref{pentaQ}. Since $\ell=0$, all the states have negative parity. Different states correspond to different spin combinations, in which $[~]$ indicates the combination having the lowest spin, while $\{\}$ is the one with the highest.

\begin{table}[h]
\caption{Masses  of $Qqqq\bar q$ pentaquarks, where $q=u,d$,  computed in model $\mathcal{B}$ described in Fig.~\ref{figuredqqq}. }
\begin{center}
\begin{tabular}{|c|c|c||c|c|c|}
\hline
\hline
Content & $J^P$& Mass (GeV)  & Content & $J^P$& Mass (GeV)\\
\hline
\hline
$[\{[[cq]\bar q]q\}q]$ &   $\frac{1}{2}^-$ & 3.36  & $[\{[[bq]\bar q]q\}q]$ &   $\frac{1}{2}^-$ & 6.69\\
\hline
$ [[[[cq]\bar q]q]q]$ &  $\frac{1}{2}^-$ & 3.37  & $ [[[[bq]\bar q]q]q]$ &  $\frac{1}{2}^-$ & 6.70   \\
\hline
$[\{[\{cq\}\bar q]q\}q]$ &   $\frac{1}{2}^-$ & 3.39 & $[\{[\{bq\}\bar q]q\}q]$ &   $\frac{1}{2}^-$ & 6.71 \\
\hline
$[[\{\{cq\}\bar q\}q]q]$ &   $\frac{1}{2}^-$ & 3.39 & $[[\{\{bq\}\bar q\}q]q]$ &   $\frac{1}{2}^-$ & 6.71 \\
\hline
$[[[\{cq\}\bar q]q]q]$ &   $\frac{1}{2}^-$ & 3.40 & $[[[\{bq\}\bar q]q]q]$ &   $\frac{1}{2}^-$ & 6.71 \\
\hline
$\{\{[[cq]\bar q]q\}q\}$ &   $\frac{3}{2}^-$ & 3.44 & $\{\{[[bq]\bar q]q\}q\}$ &   $\frac{3}{2}^-$ & 6.71 \\
\hline
$[\{\{\{cq\}\bar q\}q\}q]$ &   $\frac{3}{2}^-$ & 3.45 & $[\{\{\{bq\}\bar q\}q\}q]$ &   $\frac{3}{2}^-$ & 6.72 \\
\hline
$\{\{[\{cq\}\bar q]q\}q\}$ &   $\frac{3}{2}^-$ & 3.48 & $\{\{[\{bq\}\bar q]q\}q\}$ &   $\frac{3}{2}^-$ & 6.73 \\
\hline
$\{[\{\{cq\}\bar q\}q]q\}$ &   $\frac{3}{2}^-$ &  3.48 & $\{[\{\{bq\}\bar q\}q]q\}$ &   $\frac{3}{2}^-$ & 6.73  \\
\hline
$\{\{\{\{cq\}\bar q\}q\}q\}$ &   $\frac{5}{2}^-$ & 3.57 & $\{\{\{\{bq\}\bar q\}q\}q\}$ &   $\frac{5}{2}^-$ & 6.75 \\
\hline
\hline
\end{tabular}
\end{center}
\label{pentaQ}
\end{table}%

In order to compare  models $\mathcal{B}$ and  $\mathcal{A}$, we have computed again the spectra of $QQqq\bar{q}$, now assuming that the interaction among quarks works as in model $\mathcal{B}$, with results shown in Table \ref{pentaQQnew}. By looking at Tables \ref{pentaQQ} and \ref{pentaQQnew}, we find that the mass difference between spin-1/2 states is at most 50 MeV in the charm sector and 30 MeV in the bottom sector, while it is at most 10 MeV for states with spin 3/2 and 5/2 in the charm and bottom sectors.

\begin{table}[h]
\caption{Masses of $QQqq\bar q$ pentaquarks, with  $q=u,d$, computed in model $\mathcal{B}$ (see Fig.~\ref{figuredqqq}).}
\begin{center}
\begin{tabular}{|c|c|c||c|c|c|}
\hline
\hline
Content & $J^P$& Mass (GeV)  & Content & $J^P$& Mass (GeV)\\
\hline
\hline
$ [[[[cq]\bar q]q]c]$ &  $\frac{1}{2}^-$ & 4.57  & $ [[[[bq]\bar q]q]b]$ &  $\frac{1}{2}^-$ &  11.18  \\
\hline
$[[[\{cq\}\bar q]q]c]$ &   $\frac{1}{2}^-$ & 4.60 & $[[[\{bq\}\bar q]q]b]$ &   $\frac{1}{2}^-$ & 11.19\\
\hline
$[\{[[cq]\bar q]q\}c]$ &   $\frac{1}{2}^-$ &  4.61 & $[\{[[bq]\bar q]q\}b]$ &   $\frac{1}{2}^-$ & 11.19\\
\hline
$[[\{\{cq\}\bar q\}q]c]$ &   $\frac{1}{2}^-$ & 4.63 & $[[\{\{bq\}\bar q\}q]b]$ &   $\frac{1}{2}^-$ & 11.20 \\
\hline
$[\{[\{cq\}\bar q]q\}c]$ &   $\frac{1}{2}^-$ & 4.64 & $[\{[\{bq\}\bar q]q\}b]$ &   $\frac{1}{2}^-$ & 11.20 \\
\hline
$\{\{[[cq]\bar q]q\}c\}$ &   $\frac{3}{2}^-$ & 4.63 & $\{\{[[bq]\bar q]q\}b\}$ &   $\frac{3}{2}^-$ & 11.19 \\
\hline
$\{\{[\{cq\}\bar q]q\}c\}$ &   $\frac{3}{2}^-$ & 4.66 & $\{\{[\{bq\}\bar q]q\}b\}$ &   $\frac{3}{2}^-$ & 11.20 \\
\hline
$\{[\{\{cq\}\bar q\}q]c\}$ &   $\frac{3}{2}^-$ & 4.66 & $\{[\{\{bq\}\bar q\}q]b\}$ &   $\frac{3}{2}^-$ & 11.20  \\
\hline
$[\{\{\{cq\}\bar q\}q\}c]$ &   $\frac{3}{2}^-$ & 4.72 & $[\{\{\{bq\}\bar q\}q\}b]$ &   $\frac{3}{2}^-$ & 11.22 \\
\hline
$\{\{\{\{cq\}\bar q\}q\}c\}$ &   $\frac{5}{2}^-$ & 4.74 & $\{\{\{\{bq\}\bar q\}q\}b\}$ &   $\frac{5}{2}^-$ & 11.23 \\
\hline
\hline
\end{tabular}
\end{center}
\label{pentaQQnew}
\end{table}%

\subsection{$Qqqq\bar{Q}$}\label{sec:QQbar}
Model $\mathcal{B}$ can be used to study $Qqqq\bar{Q}$ states, in order to compare the outcomes with the masses of the states observed at LHCb. The values of the masses are shown in Table \ref{pentaQQbar}.
\begin{table}[h]
\caption{Masses (GeV) of $Qqqq\bar Q$ pentaquarks, with $q=u,d$, computed in model $\mathcal{B}$ of Fig.~\ref{figuredqqq}.}
\begin{center}
\begin{tabular}{|c|c|c||c|c|c|}
\hline
\hline
Content & $J^P$& Mass (GeV)  & Content & $J^P$& Mass (GeV)\\
\hline
\hline
$ [[[[cq]\bar c]q]q]$ &  $\frac{1}{2}^-$ & 4.57  & $ [[[[bq]\bar b]q]q]$ &  $\frac{1}{2}^-$ &  11.19  \\
\hline
$[\{[[cq]\bar c]q\}q]$ &   $\frac{1}{2}^-$ & 4.57  & $[\{[[bq]\bar b]q\}q]$ &   $\frac{1}{2}^-$ & 11.19\\
\hline
$[[\{\{cq\}\bar c\}q]q]$ &   $\frac{1}{2}^-$ & 4.58 & $[[\{\{bq\}\bar b\}q]q]$ &   $\frac{1}{2}^-$ & 11.21 \\
\hline
$[\{[\{cq\}\bar c]q\}q]$ &   $\frac{1}{2}^-$ & 4.64 & $[\{[\{bq\}\bar b]q\}q]$ &   $\frac{1}{2}^-$ & 11.22 \\
\hline
$[[[\{cq\}\bar c]q]q]$ &   $\frac{1}{2}^-$ & 4.65 & $[[[\{bq\}\bar b]q]q]$ &   $\frac{1}{2}^-$ & 11.22 \\
\hline
$\{\{[[cq]\bar c]q\}q\}$ &   $\frac{3}{2}^-$ & 4.64 & $\{\{[[bq]\bar b]q\}q\}$ &   $\frac{3}{2}^-$ & 11.20 \\
\hline
$[\{\{\{cq\}\bar c\}q\}q]$ &   $\frac{3}{2}^-$ & 4.66 & $[\{\{\{bq\}\bar b\}q\}q]$ &   $\frac{3}{2}^-$ & 11.22 \\
\hline
$\{[\{\{cq\}\bar c\}q]q\}$ &   $\frac{3}{2}^-$ & 4.67 & $\{[\{\{bq\}\bar b\}q]q\}$ &   $\frac{3}{2}^-$ &  11.22 \\
\hline
$\{\{[\{cq\}\bar c]q\}q\}$ &   $\frac{3}{2}^-$ & 4.71 & $\{\{[\{bq\}\bar b]q\}q\}$ &   $\frac{3}{2}^-$ & 11.23 \\
\hline
$\{\{\{\{cq\}\bar c\}q\}q\}$ &   $\frac{5}{2}^-$ & 4.76 & $\{\{\{\{bq\}\bar b\}q\}q\}$ &   $\frac{5}{2}^-$ & 11.24 \\
\hline
\hline
\end{tabular}
\end{center}
\label{pentaQQbar}
\end{table}%
The states with hidden charm and spin 1/2 have masses in the range $4.57-4.65$ GeV, spin-3/2 states have masses in the range $4.64-4.71$ GeV, the spin-5/2 one has mass 4.76 GeV. The LHCb Collaboration has argued \cite{Aaij:2015tga} that the first observed pentaquarks $P_c(4380)^+$ and $P_c(4450)^+$ have masses $4380\pm 8\pm 29$ MeV and $4449.8\pm 1.7\pm 2.5$ MeV, respectively, and opposite parity, while subsequent analyses \cite{Aaij:2019vzc} have shown that in this mass region there are three resonances, $P_c(4312)^+$, $P_c(4440)^+$ and $P_c(4457)^+$, with mass $4311.9\pm 0.7^{+6.8}_{-0.6}$ MeV, $4440.3\pm 1.3^{+4.1}_{-4.7}$ MeV, and $4457.3\pm 0.6^{+4.1}_{-1.7}$ MeV, respectively. 
We can try to compare these data with our theoretical predictions. The mass differences between theoretical and experimental results are equal to $260-340$ MeV for $P_c(4312)^+$ (assuming it has spin-parity $1/2^-$ \cite{newlhcb}), $130-210$ MeV for $P_c(4440)^+$ (assuming it has spin-parity $1/2^-$ \cite{newlhcb}), and  $180-250$ MeV for $P_c(4457)^+$ (assuming it has spin-parity $3/2^-$ \cite{newlhcb}). Regarding the previously observed state $P_c(4450)^+$, if we assume it has spin-parity $5/2^-$, its mass is different from the theoretical result by $310$ MeV. Finally, the discrepancy between the previously observed state $P_c(4380)^+$ and masses in Table \ref{pentaQQbar} of spin 3/2 pentaquarks is in the range $260-330$ MeV.
 Therefore, the present study suggests that the new experimental results for $P_c(4440)^+$ and $P_c(4457)^+$  are more compatible with a pentaquark spectrum with the predicted spin-parity assignment. If we estimate the theoretical error  $\lesssim 80$~MeV, as the one found when studying meson spectra with the same model and set of parameters \cite{Carlucci:2007um}, the  masses predicted in the first LHCb paper \cite{Aaij:2015tga} and the mass of $P_c(4312)^+$ \cite{Aaij:2019vzc} are significantly lower than the theoretical ones, while the newly observed states $P_c(4440)^+$ and $P_c(4457)^+$ \cite{Aaij:2019vzc}  get a better comparison, even though their masses are systematically lower.

\subsection{More on $QQqq\bar{q}$}\label{sec:QQbis}
\begin{figure}[h!]
\begin{center}
    \includegraphics[width=11.5cm]{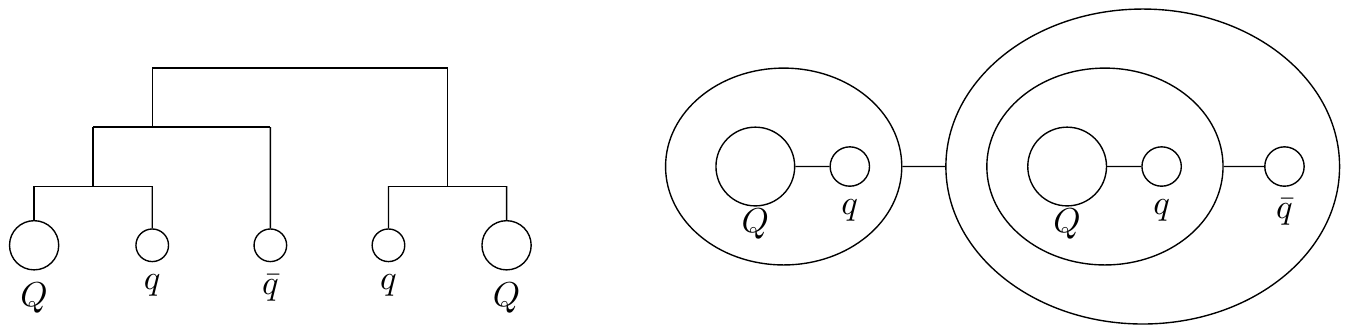}
\caption{Dendrogram and sketch of the quark content of the pentaquark in the model described in Section \ref{sec:QQbis}.}\label{figuredt}
\end{center}
\end{figure}

Within this framework, i.e. finding pentaquark masses by a sequence of two-body interactions involving at least one heavy particle, another configuration for $QQqq\bar{q}$ states is allowed. Indeed, one can consider the interaction between a $Qq$ diquark and a $Qq\bar q$ state. A similar configuration has been studied in \cite{Lebed:2015tna}-\cite{Zhu:2015bba} to compute pentaquark masses in the hidden charm sector, comprising a $cq$ diquark interacting with the triquark $\bar c (qq)$. The spirit of the computation is similar to what has been done in previous Sections, and is explained by Fig.~\ref{figuredt}.  The results for the pentaquark masses are  in Table \ref{pentaQQnewnew}. A comparison among Tables  \ref{pentaQQ}-\ref{pentaQQnew}-\ref{pentaQQnewnew} shows that the masses found in this configuration are larger than the ones found in models $\mathcal{A}$ and $\mathcal{B}$.

\begin{table}[h]
\caption{Masses of $QQqq\bar q$ pentaquarks in the diquark-triquark configuration, where $q=u,d$. }
\begin{center}
\begin{tabular}{|c|c|c||c|c|c|}
\hline
\hline
Content & $J^P$& Mass (GeV)  & Content & $J^P$& Mass (GeV)\\
\hline
\hline
$ [cq] [[cq]\bar q]$ & $\frac{1}{2}^-$ & 4.59  & $ [bq] [[bq]\bar q]$ &  $\frac{1}{2}^-$ &  11.20  \\
\hline
$[cq] [\{cq\}\bar q]$ &   $\frac{1}{2}^-$ &  4.62 & $[bq] [\{bq\}\bar q]$ &   $\frac{1}{2}^-$ & 11.21\\
\hline
$\{cq\}[[cq]\bar q] $ &   $\frac{1}{2}^-$ & 4.68 & $\{bq\}[[bq]\bar q]$ &   $\frac{1}{2}^-$ & 11.23 \\
\hline
$\{cq\} [\{cq\}\bar q]$ &   $\frac{1}{2}^-$ & 4.71 & $\{bq\} [\{bq\}\bar q]$ &   $\frac{1}{2}^-$ & 11.25 \\
\hline
$\{cq\} \{\{cq\}\bar q\}$ &   $\frac{1}{2}^-$ & 4.77 & $\{bq\} \{\{bq\}\bar q\}$ &   $\frac{1}{2}^-$ & 11.26  \\
\hline
$\{cq\}[[cq]\bar q] $ &   $\frac{3}{2}^-$ & 4.69 & $\{bq\}[[bq]\bar q]$ &   $\frac{3}{2}^-$ & 11.23 \\
\hline
$[cq] \{\{cq\}\bar q\}$ &   $\frac{3}{2}^-$ & 4.70 & $[bq] \{\{bq\}\bar q\}$ &   $\frac{3}{2}^-$ & 11.23 \\
\hline
$\{cq\} [\{cq\}\bar q]$ &   $\frac{3}{2}^-$ & 4.72 & $\{bq\} [\{bq\}\bar q]$ &   $\frac{3}{2}^-$ & 11.25 \\
\hline
$\{cq\} \{\{cq\}\bar q\}$ &   $\frac{3}{2}^-$ & 4.78 & $\{bq\} \{\{bq\}\bar q\}$ &   $\frac{3}{2}^-$ & 11.26 \\
\hline
$\{cq\} \{\{cq\}\bar q\}$ &   $\frac{5}{2}^-$ & 4.80 & $\{bq\} \{\{bq\}\bar q\}$ &   $\frac{5}{2}^-$ & 11.27 \\
\hline
\hline
\end{tabular}
\end{center}
\label{pentaQQnewnew}
\end{table}%

\section{Conclusions}\label{sec:conclusions}
We have computed pentaquark masses in a potential model. We have exploited a relativistic wave equation describing the interaction between two states, and tried to accommodate pentaquarks in this framework by considering them as emerging from three subsequent interactions, as shown in Fig.~\ref{figuraddq}, in the diquark-diquark-antiquark picture. In particular, the scheme has been used for pentaquarks with two heavy quarks ($QQ^\prime qq\bar q$). Then, we have studied $Q qqq\bar q$ pentaquarks introducing a different scheme of interaction, depicted in Fig.~\ref{figuredqqq}. The model has also been applied to $Q\bar Q qqq$ pentaquarks, with the same quark content as the states $P_c(4380)^+$ and $P_c(4450)^+$  observed  by LHCb in the hidden-charm sector, recently updated to $P_c(4312)^+$ , $P_c(4440)^+$  and   $P_c(4457)^+$. 
Comparing the predicted mass of $cqqq\bar c$ states with the experimental ones, we have found values higher than those measured by LHCb, but with a better agreement for the newly observed $P_c(4440)^+$  and   $P_c(4457)^+$ states. Further investigations could help to shed light on this discrepancy, and clarify if it is due to the approximations involved in the model. As a future study, it would be interesting to improve this potential model, making it more suitable for the description of exotic states, for instance by considering known masses of tetraquarks or pentaquarks as inputs when fixing the parameters, or by improving the choice of the potential of interaction, going beyond the one-gluon-exchange approximation.   
A possible modification in this direction can consist in introducing, for pentaquark spectroscopy, a new value for the constant term $V_0$ of the quark-antiquark potential, different from the one found when studying meson spectra. Indeed, it has been stated that constituent quark masses in potential models can get different values in baryon and meson spectroscopy \cite{Maiani:2004vq}. This discrepancy can be taken into account in this model by using a different offset for the potential, so a new value for $V_0$. If we assume that the mass of the lightest spin-1/2 pentaquark with hidden charm is 4312 MeV, we find $V_0=-0.594$ GeV. Using this value for studying the other states in the hidden-charm sector with $\ell=0$ and negative parity, we find that the mass of the heaviest spin-1/2 pentaquark is 4.39 GeV and the mass of the  heaviest spin-3/2 pentaquark is 4.45 GeV, getting a better agreement with experimental observations. Notice that spin splittings are not modified by a change in $V_0$, while a different value of quark masses could, in principle, also affect the spin-spin interaction, so a proper investigation of this aspect would deserve a dedicated study based on baryon spectroscopy.

\vspace*{1cm} {\bf \noindent Acknowledgments.}
I would like to thank   P.~Colangelo and F.~De Fazio for  valuable suggestions and discussions.

\end{document}